\documentclass[aps,prl,twocolumn,showpacs,superscriptaddress]{revtex4}
\usepackage{amsmath}
\usepackage{dcolumn}
\usepackage{epsfig}
\usepackage{graphicx}
\usepackage{latexsym}
\usepackage{amssymb}
\def\S2RO4{Sr$_2$RuO$_{4}$}
\def\SRO3{SrRuO$_{3}$}
\def\STO3{SrTiO$_{3}$}

\begin{document}

\title{Inverse proximity effect in superconductor-ferromagnet bilayer structures }

\author{Jing Xia}
\affiliation{Geballe Laboratory for Advanced Materials, Stanford University, Stanford, California, 94305}
\affiliation{Department of Physics, Stanford University, Stanford, CA 94305} 
\author{V. Shelukhin}
\affiliation{School of Physics and Astronomy, Tel Aviv University, Tel Aviv 69978, Israel}
\author{M. Karpovski}
\affiliation{School of Physics and Astronomy, Tel Aviv University, Tel Aviv 69978, Israel}
\author{A. Kapitulnik}
\affiliation{Department of Physics, Stanford University, Stanford, CA 94305}
\affiliation{Department of Applied Physics, Stanford University, Stanford, CA 94305} 
\author{A. Palevski}
\affiliation{School of Physics and Astronomy, Tel Aviv University, Tel Aviv 69978, Israel}

\begin{abstract}
Measurements of the polar Kerr effect using a zero-area-loop Sagnac magnetometer on Pb/Ni and Al/(Co-Pd) proximity-effect bilayers show unambiguous evidence for the ``inverse proximity effect," in which the ferromagnet (F) induces a finite magnetization in the superconducting (S) layer.  To avoid probing the magnetic effects in the ferromagnet, the superconducting layer was prepared much thicker than the light's optical penetration depth. The sign and size of the effect, as well as its temperature dependence agree with recent predictions by Bergeret {\it et al.}\cite{bergeret5}.
\end{abstract}


\maketitle

Recent focus in the study of proximity effect - - that is, the mutual influence of a superconductor and another electronic system in contact with it,  has been on superconductors (S) and itinerant ferromagnets (F). Because the ferromagnet is inherently spin-polarized, singlet pairs from the superconductor will only penetrate a very short distance  into the ferromagnet \cite{bergeret1,buzdin1} in a S/F bilayer. This distance is a few nanometers for strong ferromagnets such as Ni, Co or Fe,  and is a few tens of nanometer for weaker ferromagnets such as NiCu alloys.  Some spectacular effects arise when a superconductor is sandwiched between two ferromagnets in a variety of S/F/S structures. For homogeneously magnetized ferromagnetic layers, periodic $\pi$-phase shifts across the junction as a function of the thickness of the ferromagnetic layer $d_F$ were predicted  \cite{buzdin2,bergeret2}.    These will result in an oscillatory Josephson critical current, an effect that was subsequently confirmed experimentally \cite{ryazanov,kontos,blum,robinson}. For non-homogeneously magnetized ferromagnets, it was proposed that an odd-triplet component will be generated \cite{bergeret3,kadigrobov}, thus allowing the Josephson coupling to extend over larger distances, limited only by the temperature length. Experimental evidence for this effect was reported recently  \cite{sosnin,keizer}.

While much of the work on S/F proximity effect focused on the penetration of the superconducting order-parameter into the ferromagnet, very little was done to understand the penetration of the ferromagnetic order-parameter, i.e. the uniform magnetization, into the superconductor. For example, in the case of induced triplet component, a novel proximity effect will result  from the zero spin projection \cite{bergeret4,kharitonov}. The theory in this case predicts an induced magnetization in the superconductor which can vary between states that either fully screen \cite{bergeret5}  or anti-screen \cite{bergeret4}  the magnetization of the ferromagnet \cite{kharitonov}, depending on the microscopic parameters of the system. The experimental observation of this, so-called "inverse proximity effect" has been viewed as a grand challenge of the field as it would provide an important complementary confirmation of the possible triplet pairing in S/F structures.

\begin{figure}[h]
\begin{center}
\includegraphics[width=1.0 \columnwidth]{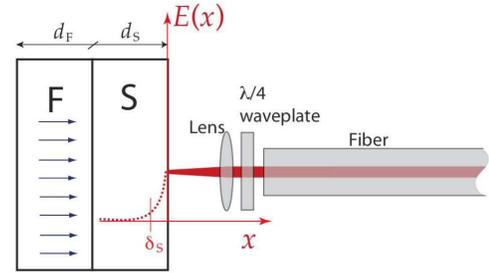}
\end{center}
\caption{ Cartoon of the measurement scheme. Here each of the two perpendicular  linearly-polarized light that comes out of the fiber \cite{xia1} becomes circularly polarized going through the  $\lambda$/4 plate and is then focused on the sample using a lens. The electric field of the incident light (represented with dashed line) penetrates a distance $\delta_S \ll d_S$ into the superconductor, thus is not sensitive to the spins in the F layer.  } 
\label{schem}
\end{figure}

In this Letter we present direct experimental observations of the inverse proximity effect in Al/(Co-Pd) and Pb/Ni bilayers.  To show unambiguously that we detect a finite magnetization signal in the superconducting layer of a S/F bilayer structure, we performed magneto-optical Kerr effect measurements using light with optical-penetration depth that is much smaller than the thickness of the superconducting layer, thus ensuring that the incident light does not interact with the moments in the ferromagnetic layer (see Fig.~\ref{schem}). We measure a finite signal that seems to onset near the superconducting transition temperature, $T_c$, of the S/F bilayer system and increases with decreasing temperature. For the Pb/Ni system, for which the superconducting coherence-length $\xi_S$, is about a half of the thickness of the superconducting film, the size of the effect is very small, of order 150 nanorad of optical rotation. For the Al/(Co-Pd) system the effect is much larger and increases in size as the temperature is lowered in accordance with predictions by Bergeret {\it et al.} \cite{bergeret5}.

Thin films of Ni deposited onto Ge or Co-Pd superlattices deposited onto Si substrates were used as the bottom ferromagnetic layers on which either Pb or Al were deposited with various thicknesses. Metals were deposited at room temperature using an electron-gun system at a vacuum of better than 1$\times$10$^{-6}$ torr.  Nickel and  superlattices of Co-Pd  were used because both these systems can be magnetized in the direction normal to the plane of the substrate with large coercive fields.   The thicknesses of the Ni layer (11 nm) or the Co-Pd superlattice (consisting of 10 periods of 0.2 nm of Co and 0.9 nm of Pd per period) were optimized to ensure the perpendicular magnetization of the films when cooled in an appropriate field from room temperature down to low temperatures. Pb and Al were chosen as two examples of long coherence length and weak spin-orbit interaction (Al) and short coherence length and strong spin-orbit interaction (Pb) superconductors. In this paper we show results on a Pb/Ni system with Pb thickness of 95 nm, and two Al/(Co-Pd) bilayer systems with Al thickness of 50 nm and 90 nm.

\begin{figure}[h]
\begin{center}
\includegraphics[width=1.0 \columnwidth]{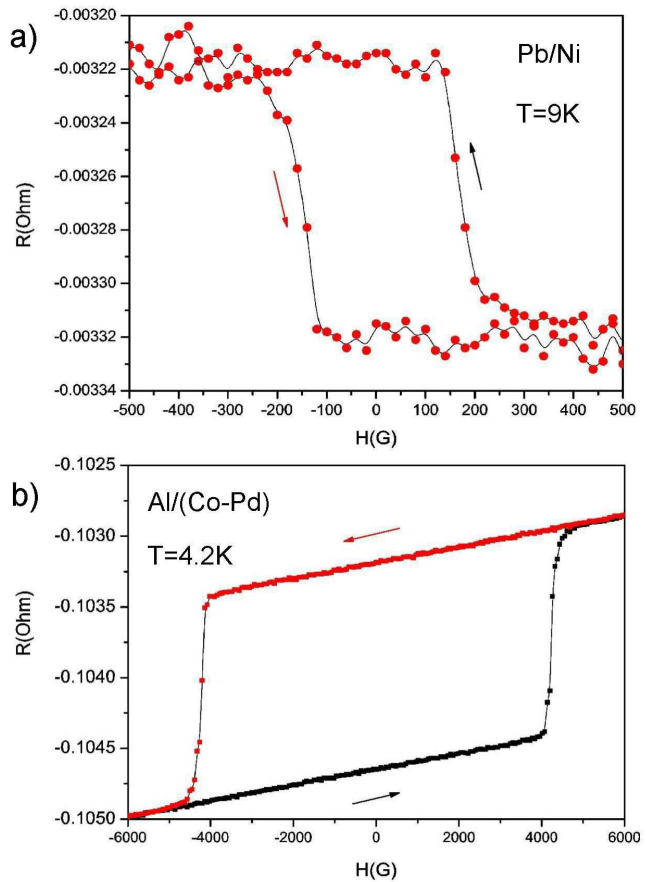}
\end{center}
\caption{ Hysteresis loops for the two systems examined measured through anomalous Hall effect. Note the sharp transitions, indicative of a moment perpendicular to the plane of current flow.} 
\label{hyst}
\end{figure}

Polar Kerr effect measurements were performed using a  zero-area-loop Sagnac interferometer at a wavelength of $\lambda =$1550 nm  \cite{xia1}.  The same apparatus was previously used to detect  time-reversal-symmetry-breaking effect below $T_c$ in Sr$_2$RuO$_4$ \cite{xia2}. Typical performance was a shot-noise limited 0.1 $\mu$rad/$\sqrt{Hz}$ at 10 $\mu$W of incident optical power from room temperature down to 0.5 K.   Samples were mounted on a copper plate using GE varnish. The system was aligned at room temperature, focusing the beam that emerges out of the quarter-waveplate to a $\sim 3 \mu$m size spot \cite{xia2}. In its used configuration, the apparatus was sensitive to only the polar Kerr effect, hence to any ferromagnetically aligned moment perpendicular to the plane of incidence of the light. Saturation Kerr angles of $\sim$170 $\mu$rad for Ni and $\sim$1.5 millirad for Co-Pd films (see also \cite{zhou}), corresponding to saturation magnetization normal to the plane of the films, were measured at low temperaturres. However, we note that the Kerr effect vanishes at around room temperature, much below the Curie temperature of either of these systems,  presumably due to reorientation transition \cite{gerber1,gerber2}.  Fig.~\ref{hyst} shows anomalous Hall effect measurements at low temperatures (but above $T_c$ of the superconductor)  which were used to determine the direction of the magnetic moment and the coercive fields in the ferromagnetic layers.  It is well established  \cite{hurd} that the magnetization perpendicular to the plane of the ferromagnet film gives rise to anomalous Hall effect. The Hall voltage is proportional to the magnetization and the proportionality coefficient depends on the material parameters.  A sharp hysteretic loop shown in Fig.~\ref{hyst} is a clear evidence of the fact that the easy axis of both ferromagnets is normal to the film's plane. A coercive field $\sim$ 200 G for Pb/Ni bilayer and $\sim$ 4 kG for Al/(Co-Pd) structure provide remnant magnetization lasting for long time. It has been checked that the magnetization perpendicular to the plane of the films at low temperatures is not changing for days at zero external magnetic field.

\begin{figure}[h]
\begin{center}
\includegraphics[width=1.0 \columnwidth]{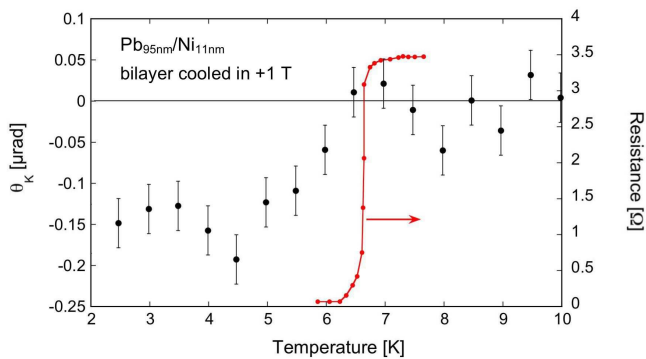}
\end{center}
\caption{ Kerr effect measurement of the Pb/Ni bilayer system. Sample was first cooled in a +1 T field down to 10K. The field was then turn down to zero, and the sample was cooled further to 0.3K. Data was taken when the sample warmed up. Also shown is the resistive transition. Note that the Kerr response indicate a magnetization that opposes the magnetization of the ferromagnetic layer.} 
\label{lead}
\end{figure}

\begin{figure}[h]
\begin{center}
\includegraphics[width=1.0 \columnwidth]{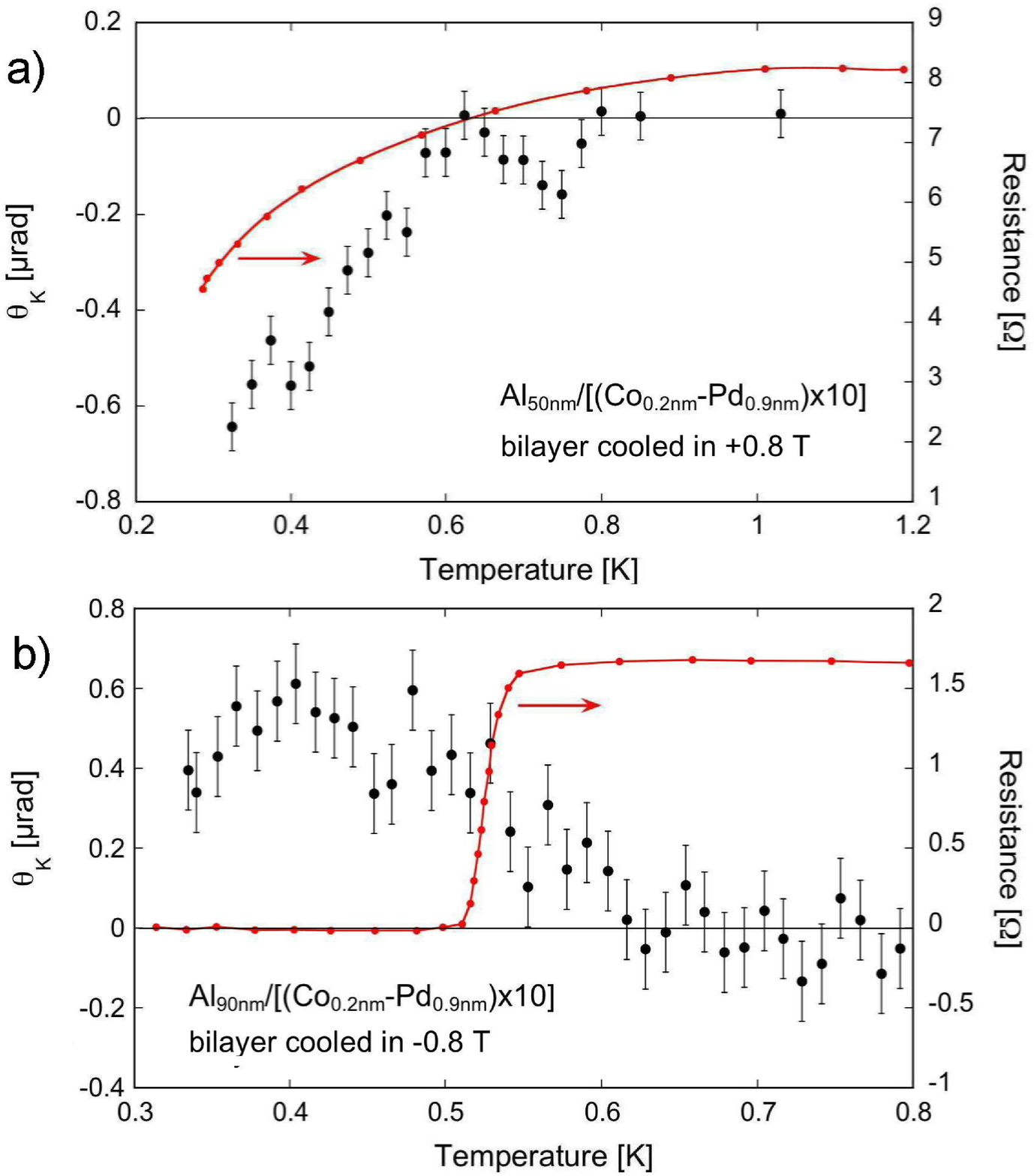}
\end{center}
\caption{ Kerr effect measurement of the Al/(Co-Pd) bilayer systems. (a) Sample with 50 nm of Al was first cooled in a field of +0.8 T down to 10 K. The field was then turn down to zero, and the sample was cooled further to 0.3K. Data was taken when the sample warmed up.  (b) Sample with 90 nm of Al was first cooled in a field of -0.8 T down to 10 K. The field was then turn down to zero, and the sample was cooled further to 0.3K. Data was taken when the sample warmed up. For both samples we also show the resistive transition. Note that the Kerr response indicate a magnetization that opposes the magnetization of the ferromagnetic layer. } 
\label{aluminum}
\end{figure}

Transport measurements were used to determine the ferromagnetic and superconducting properties of the films. Four point resistance measurements were used to determine the superconducting transition temperature ($T_c$). Fig.~\ref{lead} and  Fig.~\ref{aluminum} show the resistive transition of the bilayers studied. The zero resistance state indicating a full superconducting transition is clearly observed  for the Pb/Ni and the thicker Al in the Al/(Co-Pd) case. However, as seen in Fig.~\ref{aluminum}a, the resistive transition of the thin (50 nm) Al bilayer system is very broad and does not reach zero resistance down to 0.3 K.  

To measure the inverse proximity effect, the samples had first to be prepared with the magnetic moment perpendicular to the plane of incident light. Two procedures were used with identical results. In the first  (applied to the Pb/Ni), samples were cooled in a large applied field (0.8 to 1 tesla)  down to 12 K at which temperature the field was removed. Alternatively, samples were first cooled at zero external fields to about 10 K, at which temperature a large external magnetic field (0.8 to 1 tesla) normal to plane of the film was applied for 10 minutes to magnetize the ferromagnetic layer in the direction of the applied magnetic field. The field was then turned off and the samples were cooled to the lowest temperature ($\sim$ 0.3 K) at zero magnetic field. To ensure zero field measurements, as the superconducting magnet was turned off, it was heated above its own transition temperature to eliminate trapped flux in the magnet.  Kerr effect measurements were taken while the samples were warmed from 0.3 K (2 K in the case of Pb) to temperatures exceeding their respective superconducting $T_c$.  Fig.~\ref{lead} shows the Kerr effect measured on a Pb/Ni sample, while Fig.~\ref{aluminum} shows the Kerr effect measured on the two Al/(Co-Pd) samples.  In each of the three graphs, the onset of a Kerr angle variation  occurs at a temperature which according to our transport studies is very close to the superconductor transition temperature $T_c$ of the bilayer.

Samples were prepared with a definite direction the magnetization in the ferromagnetic layer  that without the superconducting layer would result in a definite sign for the Kerr effect. The unambiguous result for all samples is that the measured Kerr effect indicates a magnetization that opposes the one in the ferromagnetic layer.   While all bilayer structures show similar trends, the size of the effect and it temperature dependence seems to be different.  For the Pb/Ni case we note that the effect onsets rather sharply (as is superconductivity itself) below $T_c$, and it saturates to a  very small level of $\sim$150 nanorad at low temperatures. The Al/(Co-Pd) bilayer with 90 nm of Al shows similar temperature dependence as for the Pb/Ni system, but a much larger effect (Fig.~\ref{aluminum}b).  The Al/(Co-Pd) bilayer with 50 nm of Al (Fig.~\ref{aluminum}a) shows a more surprising result, that while the superconducting transition is very broad, as expected for the case of $d_S \ll \xi_S$ (here $d_S =50$ nm and $\xi_S \approx 300$ nm), the onset of the Kerr effect is much sharper and grows almost linearly as the temperature is lowered to $\sim T_c/2$. In addition we note a peculiar ``diamagnetic response" in the fluctuations regime just above $T_c$ (marked with arrows). 

To establish that indeed we measure the inverse proximity effect, we first note that the observed signal cannot result from a simple Meissner response to the magnetized ferromagnetic film. The observed signals are simply too large to result from screening currents. Moreover, because of the films are thin, they can at most be in a vortex state. However, vortices induced by the magnetized ferromagnetic films would result in a Kerr effect reflecting the magnetized ferromagnet, thus opposite to what we actually find experimentally. Note further that the magnetic field anywhere above the ferromagnetic film is vanishingly small due to the large aspect ratio resulting in a demagnetization factor very close to unity.

Following ref.~\cite{bergeret5}, we attempted to analyze our data.  The relevant parameter to compare with theory is the ratio $r_S \equiv \delta M_S(0)/M_F(0)$ of the low-temperatures saturation magnetization in the superconducting layer $\delta M_S(0)$ divided by the saturation magnetization of the ferromagnetic layer $M_F(0)$.  A crude estimate based on ref. \cite{bergeret5} indicates that $r_S \approx - c(d_F/\xi_S)$ where $c$ is a constant and the negative sign upfront indicates a magnetization that opposes in sign the magnetic alignment of the ferromagnet. The sign will remain unchanged as a function of either $d_F$ or $d_S$  as long as the bilayers are in the so-called diffusive regime \cite{kharitonov}.  However, the problem of analyzing our data is that when measuring the Kerr effect of either the S or the F sides of the bilayer, we are measuring different materials with different degree of spin-orbit interaction, hence with a different conversion of the saturated Kerr effect to saturated magnetization. Thus, in taking the ratio $r_S$ the conversion factors do not cancel. For the case of Pb/Ni there is another complication in which the film thickness is larger than the coherence length. The ratio that we find for this system is $\delta \theta_{K_S}(0)/\theta_{K_F} \approx 0.001$.

\begin{figure}[h]
\begin{center}
\includegraphics[width=1.0 \columnwidth]{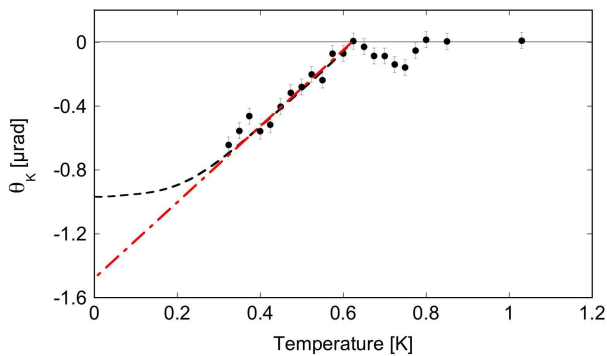}
\end{center}
\caption{ Fit of the data in Fig.~\ref{aluminum}b to the theory of Bergeret {\it et al.} \cite{bergeret5}. Dashed line and dashed-dotted lines are the two limiting cases that the data can extrapolates (see text).}
\label{fit}
\end{figure}

For the Al/(Co-Pd) system with 50 nm of Al, $d_S \ll \xi_S$ and thus the reduction of the effect due to the thickness is minimal. Here we estimate $\xi_S \approx $ 300 nm,  as the geometric mean between the bulk, disorder free coherence length for Al ($\sim 2 \mu$m) and the thickness of the film. The temperature dependence for the Kerr effect is roughly linear down to $T_c/2$  (here we determine $T_c$ as $\sim$ 0.6 K, or where the Kerr signal starts to increase) as expected from ref. \cite{bergeret5} . It is more difficult to extrapolate the temperature dependence below $T_c/2$ Thus, in Fig.~\ref{fit} we show two possible extrapolations, representing two limiting cases. The  dashed line represents a case for small $\gamma \equiv (R_b\sigma_F)/\xi_0$, while the dashed-dotted line represents the case for large $\gamma$.  Here $R_b$ is the interface resistance, $\sigma_F$ is the conductivity of the ferromagnet and $\xi_0$ is the BCS coherence length of the superconductor. While we do not have precise values for these various quantities, the believed good interface and the very long $\xi_0$ for Al may point for the dashed-line curve that saturates at low temperatures. Thus, with the above assumptions, we also obtain $\delta \theta_{K_S}(0)/\theta_{K_F} \approx 0.001$.

In conclusion we observed the inverse proximity effect in which the magnetization in a ferromagnetic film, induces a magnetization that is much smaller and opposite in sign in a superconducting film in a superconductor-ferromagnet  bilayer proximity system. This observation may lead to a more quantitative description of the general S/F proximity effect.

{\bf Acknowledgements:} Work supported by the US-Israel BSF and the Israel Ministry of Science .  Fabrication of the Sagnac system was  supported by Stanford's Center for Probing the Nanoscale, NSF NSEC Grant 0425897. Work at Stanford was supported by the Department of Energy Grant  DE-AC02-76SF00515.

\end{document}